\documentclass[12pt,a4paper]{article}
\usepackage{amssymb,amsmath,amscd,epsfig}
  \title{Spin flip of neutrinos with magnetic moment in core-collapse supernova}
\author{
 Oleg Lychkovskiy\thanks{e-mail: lychkovskiy@itep.ru}\hspace*{2mm}$^{\rm a,b}$, Sergei Blinnikov\hspace*{2mm}$^{\rm a,c}$
\\ ${\rm ^a}$ {\small\it Institute for Theoretical and Experimental Physics}\\
{\small\it 117218, B.Cheremushkinskaya 25,
Moscow, Russia}\\
${\rm ^b}$ {\small\it Moscow Institute of Physics and Technology
}\\{\small\it 141700, 9, Institutskii per., Dolgoprudny, Moscow
Region, Russia}\\
 ${\rm ^c}$ {\small\it   IPMU, University of Tokyo,
5-1-5 Kashiwanoha, Kashiwa, 277-8568, Japan}}

\date{}

\begin{document}

%\preprint{\vtop{
%{\hbox{IPMU09-0068}\vskip-0pt
%}
%}
%}

\newcommand{\nue}{\nu_e}
\newcommand{\nul}{\nu_l}
\newcommand{\antinue}{\bar{\nu}_e}
\newcommand{\ar}{\rightarrow}
\newcommand{\C}{{\rm C}}
\newcommand{\Hy}{{\rm H}}
\newcommand{\F}{{\rm F}}
\newcommand{\Co}{{\rm Co}}
\newcommand{\N}{{\rm N}}
\newcommand{\Ox}{{\rm O}}
\newcommand{\Fe}{{\rm Fe }}
\newcommand{\be}{\begin{equation}}
\newcommand{\ee}{\end{equation}}
\newcommand{\bc}{\begin{cases}}
\newcommand{\ec}{\end{cases}}
\newcommand{\B}{{\bf B}}
\newcommand{\Bp}{{\bf B_\perp}}
\newcommand{\rb}{\right)}
\newcommand{\lb}{\left(}
\newcommand{\kpc}{{\rm kpc}}

\maketitle

\hfill Preprint IPMU09-0068

\begin{abstract}
 Neutrino with magnetic moment can experience a chirality flip while scattering off charged particles. This effect may lead to important consequences for the dynamics and the neutrino signal of the core-collapse supernova. It is known that if neutrino is a Dirac fermion, then $\nu_L\ar\nu_R$ transition, induced by the chirality flip, leads to the emission of right-handed neutrinos, which are sterile (almost do not interact with matter). The typical energies of these sterile neutrinos are rather high, $E\sim(100-200)$ MeV.  Neutrino spin precession in the magnetic field either inside the collapsing star or in the interstellar space may lead to the backward transition, $\nu_R\ar\nu_L.$ Both possibilities are known to be interesting. In the former case high-energy neutrinos can deliver additional energy to the supernova envelope, which can help the supernova to explode (Dar's scenario of supernova explosion). In the latter case high-energy neutrinos may be detected simultaneously with the "normal" supernova neutrino signal, which would be a smoking gun for the Dirac neutrino magnetic moment. We report the results of the calculation of the supernova right-handed neutrino luminosity up to 250 ms after bounce, based on a dynamical model of the collapse. They allow to refine the estimates of the energy injected in the supernova envelope in the Dar's scenario. Also the sensitivity of water Cherenkov detectors to the Dirac neutrino magnetic moment is estimated. For $\mu_{\rm \nu~Dirac}=10^{-13}\mu_B$ Super-Kamiokande is expected to detect at least few high-energy events from a galactic supernova explosion.

 Also we briefly discuss the case of Majorana neutrino magnetic moment. It is pointed out that in the inner supernova core spin flips may quickly equilibrate electron neutrinos with non-electron antineutrinos if $\mu_{\rm \nu~Majorana}\gtrsim10^{-12}\mu_B.$ This may lead to various consequences for supernova physics.

\end{abstract}

%{~~~~~~~~~~~~~~~~~~~~~~~~~~~~~~~~~~~~~~~~~~~~~~~~~~~~~~~~~~~~~~~~PACS-2006: 14.60.Pq; 26.50.+x.}\\

\\
\newline

 \section{Introduction}

A straightforward way to account for neutrino masses is to introduce three singlet right-handed neutrinos (one per generation) in addition to three left-handed neutrinos of the Standard Model (SM). % which are the upper components of lepton doublets.
This allows to generate neutrino masses in the same way as up-quark masses, i.e. through the standard Higgs mechanism. Neutrinos, as quarks and charged leptons, are Dirac fermions in this case. Neutrino-Higgs vertexes are the only tree level vertexes which include right-handed neutrinos. Tiny neutrino masses imply tiny couplings with Higgs, therefore right-handed neutrinos interact extremely weakly with matter. In other words, they appear to be nearly {\it sterile}\footnote{Not to be confused with sterile left-handed neutrinos, which constitute additional lepton generations; we do not consider such neutrinos.}. In fact, measurements of the invisible Z-boson decay width (see e.g. \cite{Decamp:1989fr}) and cosmological considerations (see e.g. \cite{Barger:2003zg}) tell us that there are only three active neutrino species, which are $\nu_{e L},~\nu_{\mu L},~\nu_{\tau L}$ (along with their antiparticles), and therefore right-handed Dirac neutrinos {\it should} be nearly sterile in {\it any} extension of the SM (with any additional particles and interactions).
% as the ... data constrains the total number of active light neutrino species to be equal to 3\cite{smth}.

Neutrinos may acquire magnetic moments through the loop diagrams. In the minimal extension of the SM (only three right-handed neutrinos added) magnetic moment of the neutrino mass eigenstate $\nu_i$  is proportional to its mass $m_i$ and reads (see \cite{Lee:1977tib},\cite{Fujikawa:1980yx})

\be \label{SM magnetic moment}
\mu_i=\frac{3eG_{\rm F}m_i}{8 \sqrt{2} \pi^2 }=3.2 \cdot 10^{-19}\frac{m_i}{1~{\rm eV}} \mu_B.
\ee
This value seems to be too small to produce any observable effect. However, a number of extensions of the SM exist in which neutrino magnetic moments are orders of magnitude larger (see one of the pioneering papers \cite{Voloshin:1987qy} and a review \cite{Vysotsky:2002yu} with further references therein).

% and massive stars with $M=(7-18)M_\odot$ \cite{}

Historically a considerable interest to the possibility of large neutrino magnetic moment was caused by a proposition to explain the Solar neutrino deficit through the neutrino spin precession in the magnetic field of the Sun. The idea was first presented in 1971 \cite{Cisneros:1970nq}, and elaborated on in a set of papers \cite{Voloshin:1986ty}-\cite{Voloshin et al} in 1986-1987. Later, however, the neutrino flavor mixing was established to be a correct solution of the Solar neutrino problem; as for the spin precession hypothesis, neutrino magnetic moment values which it implied were disfavored by the astrophysical constraints (which are discussed below).

Core-collapse supernova was soon realized to be another astrophysical object for which neutrino magnetic moment could be important. In the beginning of the year 1987 A. Dar\\ proposed a scenario of supernova explosion based on the two-stage $\nu_L\ar\nu_R\ar\nu_L$ transition of Dirac neutrinos, where the first stage could occur in the supernova core due to the electromagnetic scattering of neutrinos on charged particles, and the second one -- in the supernova envelope due to the neutrino spin precession in the magnetic field of the star \cite{Dar:1987yv}. The detection of neutrinos from a nearby supernova SN1987A on February 23, 1987 triggered a bunch of papers \cite{Goldman:1987fg}-\cite{Notzold:1988kz} (see also a related paper \cite{Nussinov:1987zr}), which idea was closely related to the Dar's one. Namely, the supernova core was regarded as a source of right-handed sterile neutrinos, $\nu_R,$ as in \cite{Dar:1987yv}, but the $\nu_R\ar\nu_L$ transition was assumed to occur in the interstellar magnetic field. It was shown that this could lead to the registration of high-energy neutrino events in terrestrial detectors simultaneously with the ordinary supernova neutrino signal. The absence of such events and the estimation of the supernova core cooling rate due to the sterile neutrino emission was used to put stringent bounds on the neutrino magnetic moment (see Table \ref{bounds on mu}). Different aspects of the role of neutrino magnetic moment in the supernova explosion and neutrino emission were further investigated in \cite{Voloshin:1988xu}-\cite{Blinnikov:1988xq}. In particular, it was pointed out in \cite{Voloshin:1988xu}\cite{Voloshin:1988xs} that a {\it resonant} $\nu_R\ar\nu_L$ transition could proceed inside a supernova, which could greatly facilitate the explosion through the Dar's mechanism and under certain circumstances to annul the supernova bounds on the neutrino magnetic moment.

The neutrino magnetic moment could also play a role in the cooling of stars through the plasmon decay in two neutrinos. This was first pointed out as early as 1963 \cite{Bernstein:1963qh}; in this work first astrophysical bound on $\mu_\nu$ was derived from the cooling rate of the Sun. Later cooling of different types of stars, He burning stars (see e.g \cite{Fukugita:1987uy}), white dwarfs (see e.g. \cite{Blinnikov&Dunina}) and red giants (see e.g. \cite{Raffelt:1999gv}) in particular, was studied to put bounds on the neutrino magnetic moment. %Currently the best constraints are obtained from the cooling rates of red giants \cite{} and white dwarfs \cite{}, see Table \ref{bounds on mu}.

The neutrino magnetic moment should slightly change the neutrino-electron scattering cross-section. This fact underlies the laboratory experiments which search for the neutrino magnetic moment. Borexino \cite{Arpesella:2008mt}, GEMMA \cite{Beda:2007hf} and MUNU \cite{Daraktchieva:2003dr} experiments currently provide the best limits. The most relevant limits on the neutrino magnetic moment are summarized in Table \ref{bounds on mu}.

\begin{table}[t]\label{bounds on mu}
\begin{center}
\begin{tabular}{|l|c|}
\hline
Laboratory experiment GEMMA \cite{Beda:2007hf}& $\mu_\nu< 5.8 \cdot 10^{-11}\mu_B~~~~90\%$ CL \\
\hline
Cooling rates of white dwarfs \cite{Blinnikov&Dunina} & $\mu_\nu \lesssim  10^{-11}\mu_B$\\
\hline
Cooling rates of red giants (see e.g. \cite{Raffelt:1999gv}) & $\mu_\nu \lesssim 3\cdot10^{-12}\mu_B$\\
\hline
Supernova energy losses \cite{Kuznetsov:2009zm}& $\mu_\nu \lesssim (1.1-2.7)\cdot10^{-12} \mu_B$\\
\hline
Absence of high-energy events  & $\mu_\nu \lesssim 10^{-12} \mu_B$\\
in the SN1987A neutrino signal \cite{Barbieri:1988nh} & \\
\hline
\end{tabular}
\end{center}
\caption{Some bounds on the neutrino magnetic moment.
 %The cited direct laboratory bound is the strongest one.
 The first three bounds apply to both Dirac and Majorana magnetic moments. The last two bounds apply only to Dirac magnetic moment.
 %The cited astrophysical bounds are ranged according to the degree of their dependance on specific astrophysical models (the stronger bounds correspond to the more model-dependant considerations). See the text for more details and references.
 }
\end{table}

%One can see that the astrophysical constraints are more than one order of magnitude stronger than the laboratory ones. However, they are model-dependant and less definite.

In the present paper we calculate the supernova core luminosity in right-handed Dirac neutrinos up to 250 ms after core bounce.
%We assume that neutrinos are Dirac particles with the magnetic moment in the range $(10^{-12}$-$10^{-14})\mu_B.$
We use a dynamical supernova model in contrast with all previous studies \cite{Dar:1987yv}-\cite{Notzold:1988kz}, \cite{Kuznetsov:2007mp}, \cite{Ayala:1998qz}, \cite{Lychkovskiy:2008da}.\footnote{In a recent paper \cite{Kuznetsov:2009zm} dynamical supernova models are employed too.} Also we use an accurate expression for the spin-flip rate \cite{Kuznetsov:2007mp}\cite{Elmfors:1997tt} in contrast with the early studies \cite{Dar:1987yv}-\cite{Notzold:1988kz} and with \cite{Ayala:1998qz}\cite{Lychkovskiy:2008da}. We implement our result to refine the estimate of the energy injected in the supernova envelope in the Dar's scenario. Also we calculate the expected number of high-energy neutrino events in a water Cherenkov detector for a galactic supernova explosion.
The above-mentioned advantages of the present study allow us to
substantially diminish some of the uncertainties which existed
previously.

Although the main subject of the paper is neutrino spin flip due to the Dirac magnetic moment, we also briefly comment upon the case of  Majorana magnetic moment. We point out that in this case spin flips may effectively convert electron neutrinos to non-electron antineutrinos inside the inner supernova core. This may lead to various consequences for supernova physics.

%In this case neutrino spin flips convert  lead to substantial alteration of the neutrino content of the {\it neutronization burst} \cite{Lychkovskiy:2008da}. Namely, a considerable fraction of {\it antineutrinos} appears in addition to electron { \it neutrinos}, which constitute the neutronization burst in the standard picture of collapse. This may drastically increase the number of events from the neutronization burst in water Cherenkov detectors and serve a smoking-gun signature of Majorana magnetic moment. We present an estimate which shows that Majorana magnetic moments around $10^{-12}\mu_B$ are sufficient to produce observable effect in Super-Kamiokande.

 %then neutrino spin flips lead to substantial alteration of the neutrino content of the {\it neutronization burst}. Namely, a considerable fraction of {\it antineutrinos} appears in addition to electron { \it neutrinos}, which constitute the neutronization burst in the standard picture of collapse. This may drastically increase the number of events from the neutronization burst in water Cherenkov detectors and serve a smoking-gun signature of Majorana magnetic moment.

\section{Right-handed Dirac neutrino emission from supernova core}

\begin{figure}[p]
\centerline{\epsfig{file=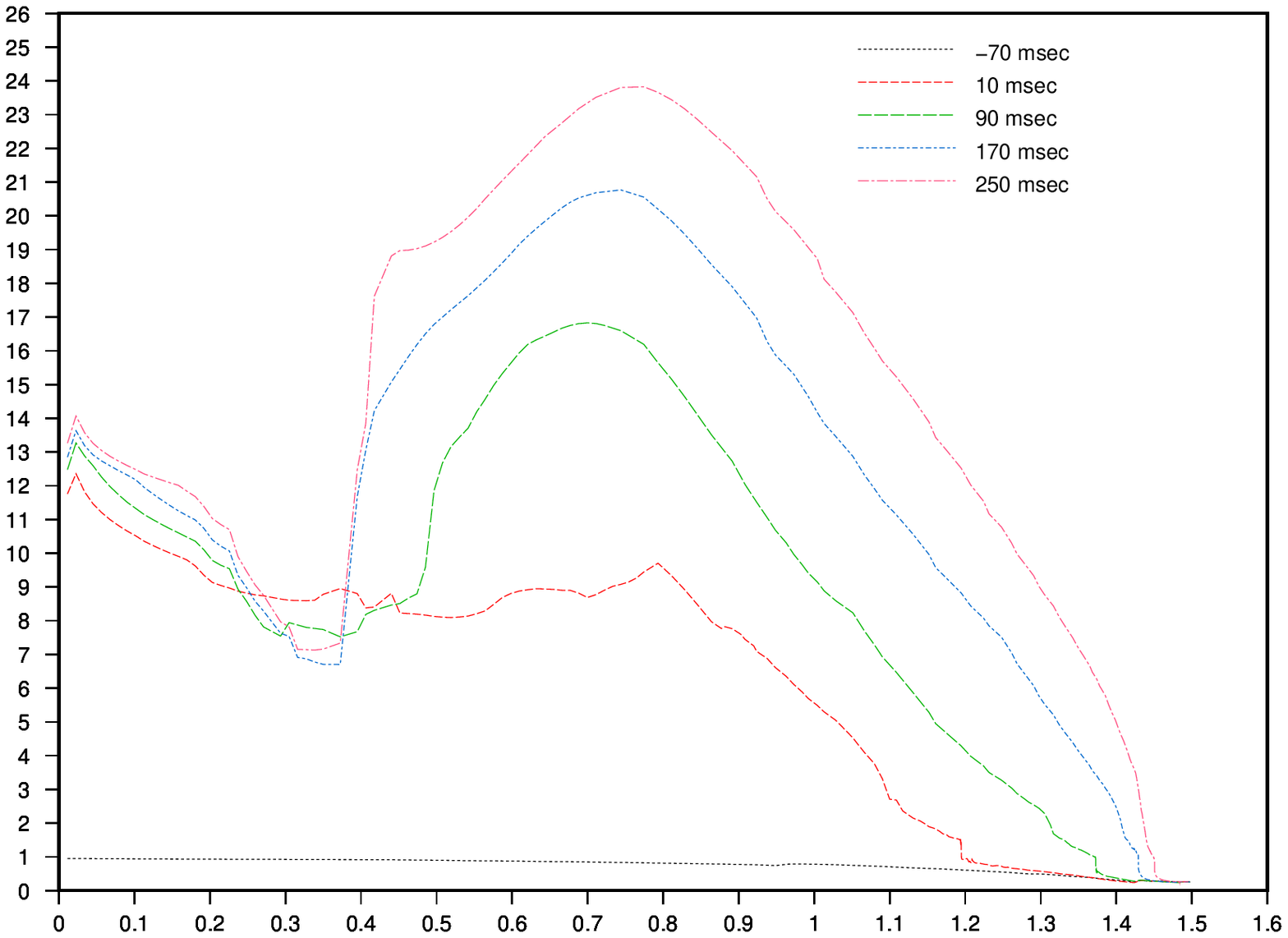, width=8.5cm}
\epsfig{file=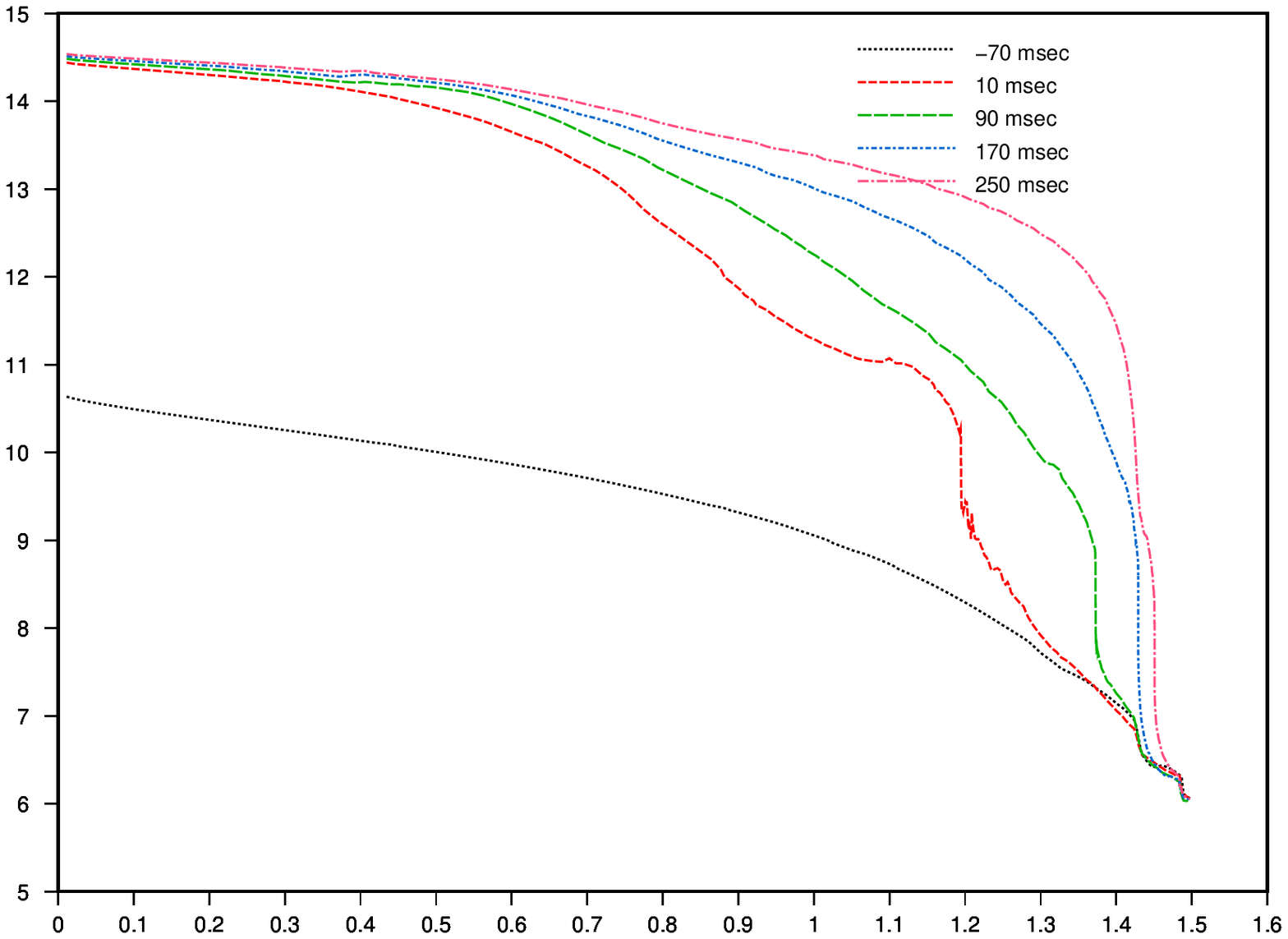, width=8.5cm}}
\caption{Left: supernova core temperature in MeVs. Right:  logarithm of supernova core density in g$/$cm$^3.$ The mass coordinate here and in what follows is measured in Sun muss units. Time here and in what follows is time after bounce. Bounce occurs at 230 ms after the beginning of the infall. }
\label{T rho}
\end{figure}

\begin{figure}[p]
\centerline{\epsfig{file=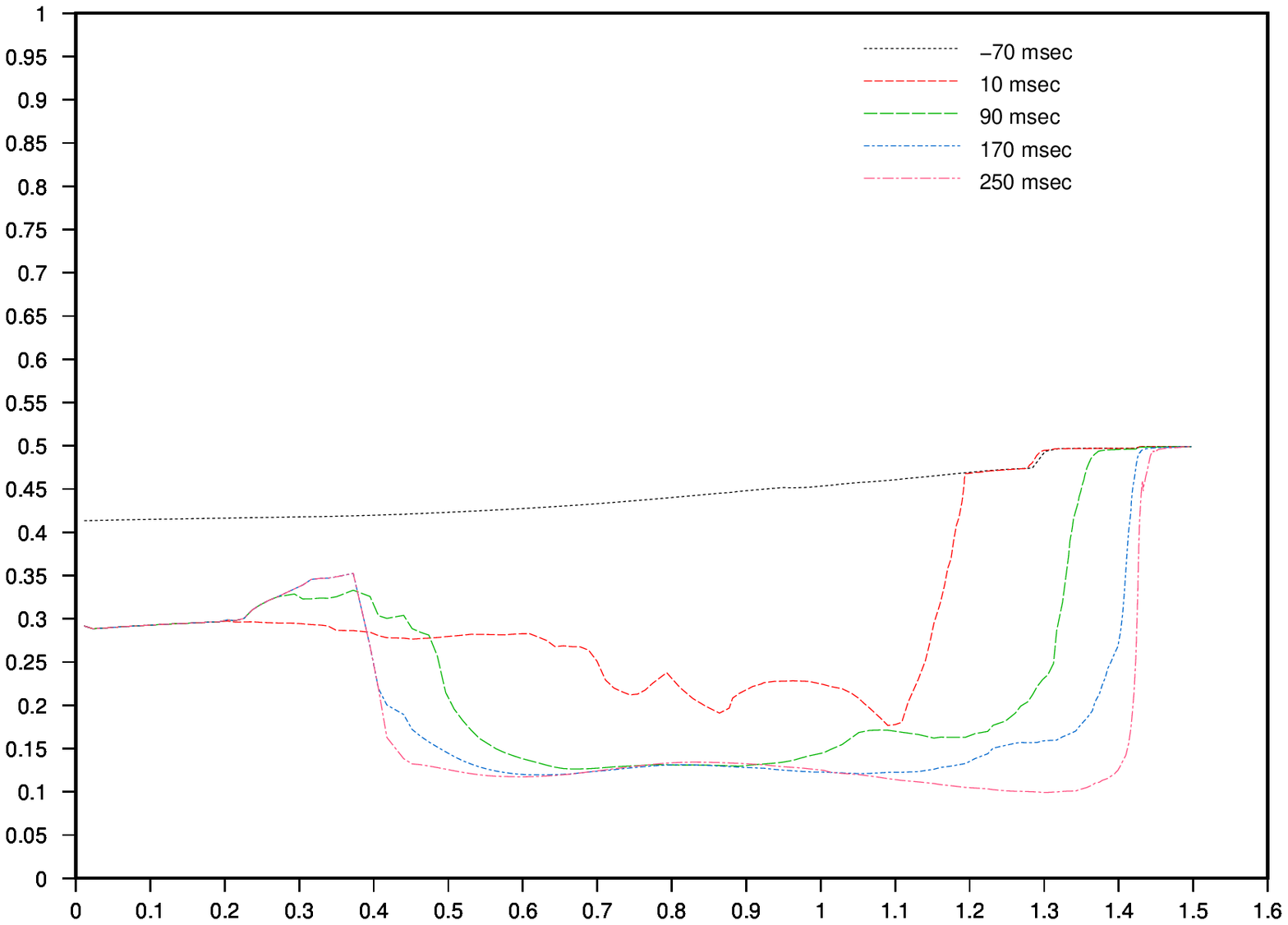, width=8.5cm}\epsfig{file=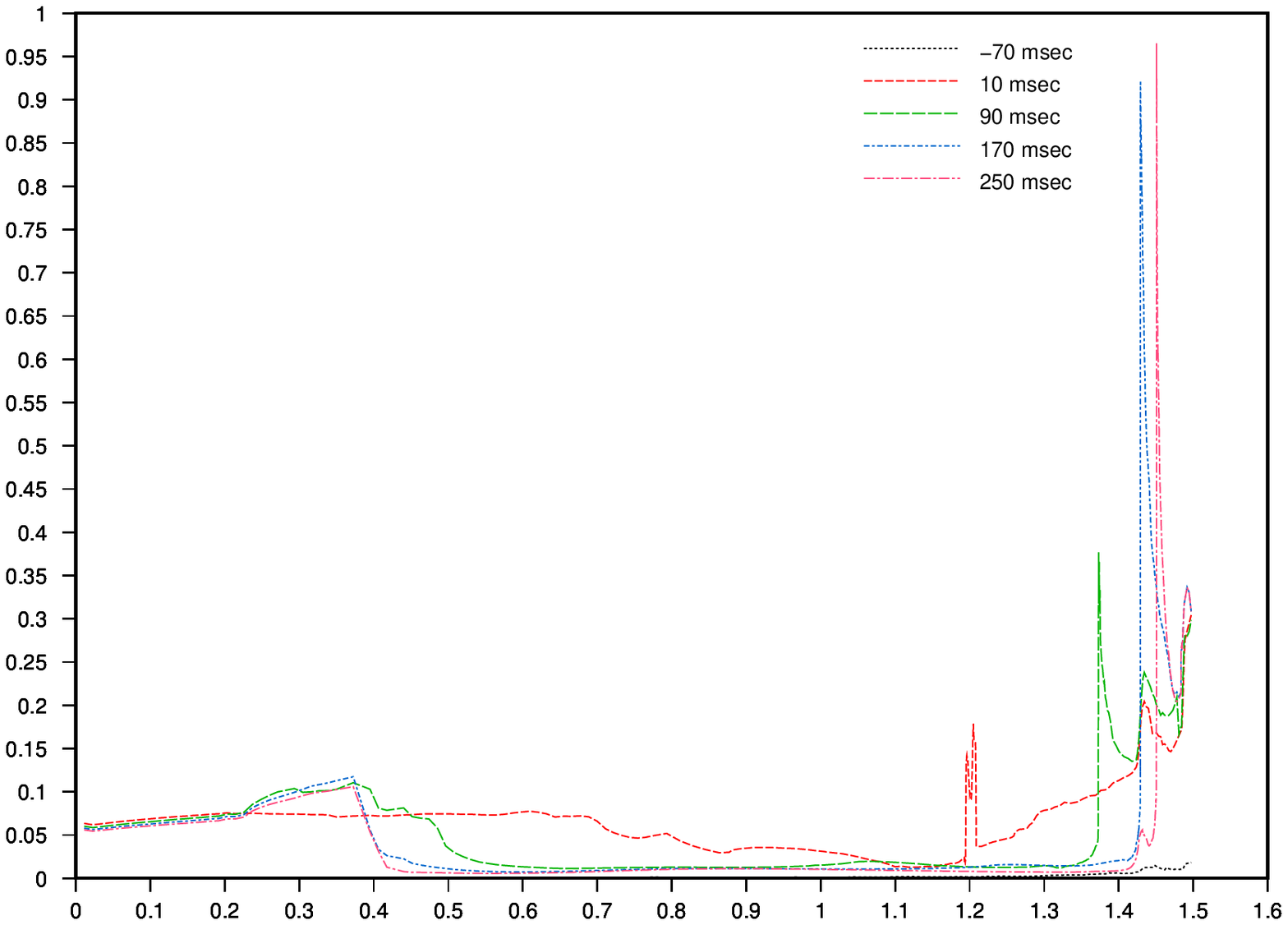, width=8.5cm}}
\caption{Electron fraction (left) and neutrino fraction (right) inside the supernova core.}
\label{ye ynu}
\end{figure}

Although numerous studies have failed to reproduce an explosion of a core-collapse supernova, there exists a commonly accepted general picture of the collapse, see e.g. review \cite{Bethe:1990mw}.  When the mass of the iron core of a massive star reaches the Chandrasekhar limit, the infall phase of the collapse starts. The core contracts due to the gravitational attraction. Some fraction of electrons is converted to electron neutrinos through the inverse beta processes. When the density of the inner part of the core reaches the nuclear density value, $\sim 3\cdot 10^{14}$ g/cm$^3,$ the infalling matter of the outer core bounces from it. A shock wave is created; it propagates outwards increasing the temperature up to tens of MeV and dissociating heavy nuclei into nucleons.

%The infall phase lasts $\sim 200$ ms; it ends exactly when density and tempret
When the densities and temperatures reach extreme values, a fraction of left-handed neutrinos experience spin flips in collisions with charged particles \cite{Dar:1987yv}. After the bounce, mainly protons, neutrons and leptons constitute the core, therefore spin flips on electrons and protons play the major role:
$$\nu_L +e \ar \nu_R +e$$
\be\label{spin flip reactions}
\nu_L +p \ar \nu_R+ p.
\ee

During a short period of time in the end of the infall (few milliseconds), when the density is high, but the temperature is low, a coherent spin flip scattering on nuclei may dominate \cite{Notzold:1988kz}. We believe that its contribution to the total  (integrated over hundreds of milliseconds after bounce) right-handed neutrino output is relatively small. Therefore we do not take it into account in the present work.
 %spin flips on nuclei while calculating core luminosity in right-handed Dirac neutrinos.
%However it should be mentioned that contribution of nuclei may be important for spin flips of {\it Majorana} neutrinos which lead to the important alteration of the neutrino content of the neutronization burst. We briefly comment on this possibility in Section 4.

%The rate of these reactions is proportional to $\mu_\nu^2.$ Sterile right-handed neutrinos escape from the core without energy loss. The spectrum of these leaking neutrinos peaks at (150-200) MeV, which is approximately equal to the neutrino  (compare this with 10-40 MeV energy range for "standard" supernova neutrinos, which are emitted from the much colder region -- neutrino sphere).
The rate of the emission of right-handed neutrinos from a supernova core reads
\be\label{emission rate}
\frac{dN_{\nu_R}}{dEdt} =  \int d^3r \frac{dn_{\nu_R}}{dEdt}(E,n_e(r,t),n_\nu(r,t), T(r,t)).
\ee
Here  ${dn_{\nu_R}}/{dEdt}$ is a spin flip rate, i.e. the number of right-handed neutrinos with energy $E$ emitted per unit energy interval per unite time from unite volume of supernova matter with temperature $T(r,t),$ electron and neutrino number densities  $n_e(r,t)$ and $n_\nu(r,t)$ correspondingly. The integration is performed over the volume of the supernova core. Note that in the first hundreds of milliseconds of the collapse only electron neutrinos are numerous inside the core; $\mu$- and $\tau$-neutrinos, as well as  antineutrinos of all flavors, are nearly absent. Therefore it is left-handed electron neutrinos which experience spin flips and turn to right-handed neutrinos. Thus $n_\nu(r,t)$ is the number density of electron neutrinos.

Two major ingredients are necessary to perform the calculation of $\frac{dN_{\nu_R}}{dEdt}.$ They are, firstly, the spin-flip rate $\frac{dn_{\nu_R}}{dEdt}$ as a function of supernova matter parameters, $T,$ $n_e$ and $n_\nu,$ and, secondly, the supernova mater parameters themselves as functions of time and coordinate, $T=T(r,t),$ $n_e=n_e(r,t)$ and $n_\nu=n_\nu(r,t).$ %Below we briefly comment on these ingredients.

%\subsection{Neutrino spin flip rate}
Neutrino spin flip is due to the exchange of a photon between a charged fermion and a neutrino with magnetic moment.\footnote{Earlier another mechanism of the neutrino spin flip in supernova was discussed. It is based on the mismatch between chirality and helicity states of massive Dirac neutrinos (see e.g. the detailed paper \cite{Burrows:1992ec} and references therein). However current upper bounds on neutrino masses ensure that this mechanism can not contribute significantly to the neutrino spin flip rate in supernova.% is irrelevant for supernova physics.
} The cross-section of these reactions is proportional to $\mu_\nu^2.$ As the process occurs in the extremely hot and dense plasma of a supernova core, photon dispersion in medium should be taken into account. Early studies \cite{Dar:1987yv}-\cite{Barbieri:1988nh} relied on the simplified expressions for the spin flip rate. An accurate expression, which is used in the present paper, was obtained in \cite{Kuznetsov:2007mp}\cite{Elmfors:1997tt}. It is rather bulky; therefore we do not quote it and refer the reader to the original papers.

\begin{figure}[t]
\centerline{\epsfig{file=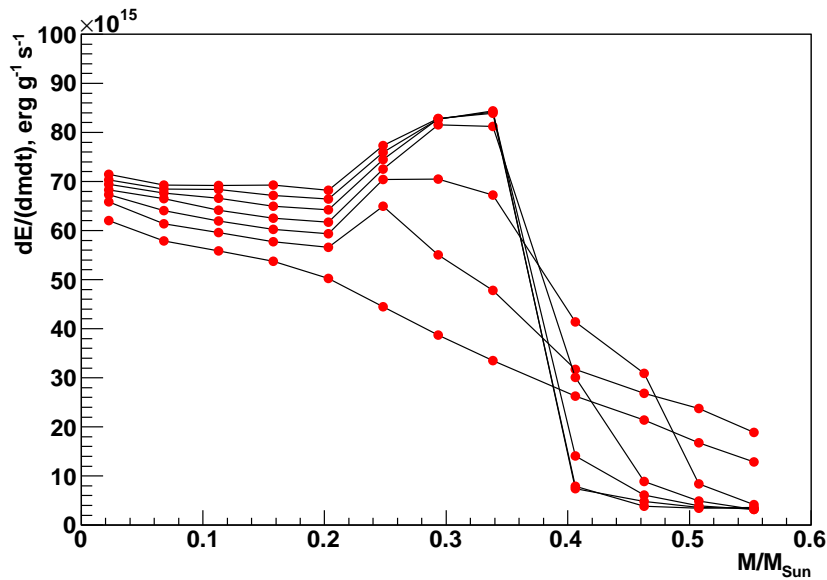, width=12 cm}}
\caption{Emittance of the supernova matter in right-handed neutrinos plotted for times 10 - 250 ms after bounce with a 40 ms step. The emittance in the center of the core monotonically grows with time. The neutrino magnetic moment is taken to be $10^{-13}\mu_B.$}
\label{emittance}
\end{figure}

\begin{figure}[t]
\centerline{\epsfig{file=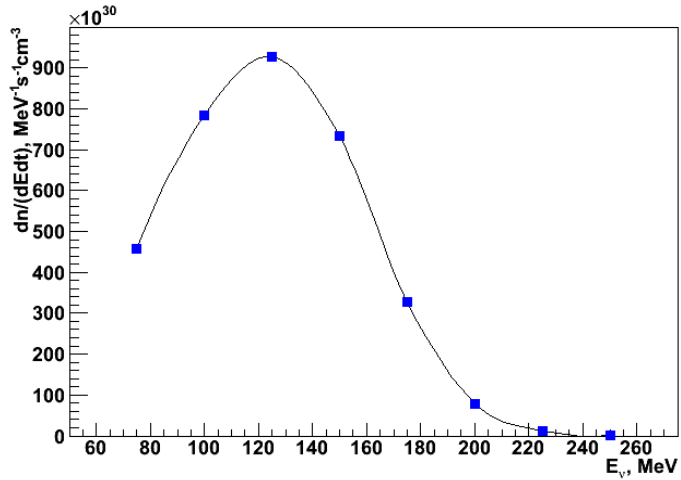, width=10cm}}
\caption{Spectrum of the emitted right-handed neutrinos for $\mu_\nu=10^{-13}\mu_B$ and for typical supernova matter parameters, $T_{\rm eff}\simeq 10$ Mev,  $\eta_{e~{\rm eff}}\simeq 250$ Mev and $\eta_{\nu~{\rm eff}}\simeq 170$ MeV.}
\label{spectrum}
\end{figure}

\begin{figure}
\centerline{\epsfig{file=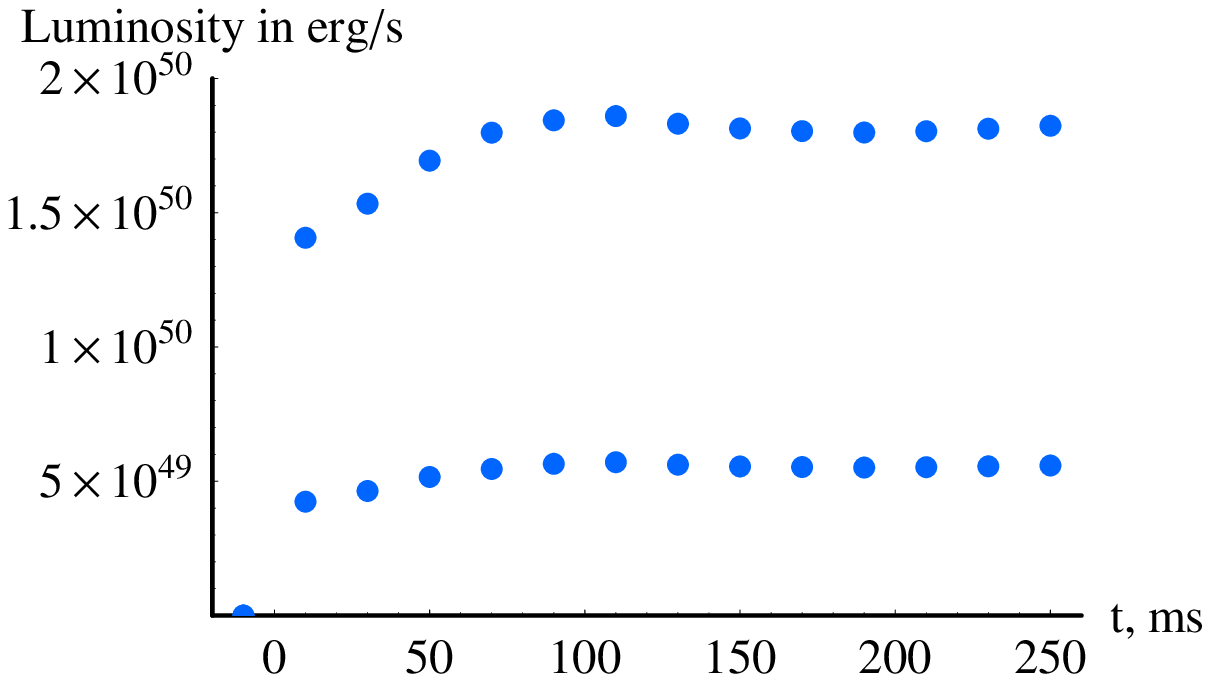, width=12 cm}}
\caption{Total luminosity of the supernova core in right-handed neutrinos for $\mu_\nu=10^{-13}\mu_B.$ The upper and the lower curves correspond to a more soft and more stiff EOS of nuclear matter accordingly.}
\label{luminosity}
\end{figure}

To obtain supernova matter state parameters as functions of coordinate and time, we employ a one-dimensional astrophysical code "Boom" \cite{Boom}. A collapse of a $1.5 M_\odot$ iron core is numerically simulated. The bounce occurs in 230 ms after the beginning of the infall. The simulation ends at 250 ms after bounce. Thus it covers almost 0.5 seconds of the collapse.
The profiles of temperature, density, electron and neutrino fractions $Y_e$ and $Y_\nu$ correspondingly are presented at Fig.\ref{T rho} and Fig.\ref{ye ynu}. The time is countered from the moment of the bounce.

As was mentioned above, only $t \gtrsim 0$ are relevant for the right-handed neutrino emission, as the supernova matter is not sufficiently dense and hot before bounce. The emittance of the supernova matter in right-handed neutrinos (energy per gram per second) is plotted at Fig.\ref{emittance} (a reference value $\mu_\nu=10^{-13}\mu_B$ is used here and in what follows). One can estimate the effective parameters of the emitting matter: $M_{\rm eff}\simeq 0.6 M_\odot,$  $T_{\rm eff}\simeq 10$ MeV, $\rho_{\rm eff}\sim 10^{14}$ g$/$cm$^3,$  $\eta_{e~{\rm eff}}\simeq 250$ MeV and $\eta_{\nu~{\rm eff}}\simeq 170$ MeV, chemical potentials $\eta_e$ and $\eta_\mu$ being related to electron and neutrino number densities by (see e.g. \cite{Blinnikov et al})
$$n_e=\frac{1}{3\pi^2}(\eta_e^3+\pi^2\eta_e T^2)$$
\be\label{Fermi energies}
n_\nu=\frac{1}{6\pi^2}(\eta_\nu^3+\pi^2\eta_\nu T^2).
\ee
These parameters are somewhat different from what was usually assumed in the previous works \cite{Barbieri:1988nh}\cite{Kuznetsov:2007mp}\cite{Ayala:1998qz}.
In particular, we emphasize that only the inner supernova core contributes significantly to the $\nu_R$ emission. The evident reasons is that both neutrino and electron chemical potentials are high inside the inner core, but fall drastically in the outer core.

The spectrum of the emitted right-handed neutrinos for the effective supernova matter parameters is plotted at Fig.\ref{spectrum}. It is peaked at $\sim 130$ MeV, which is several times larger than typical energies of ``ordinary'' neutrinos thermally emitted from the neutrino sphere.

%Let us discuss the uncertainties of the collapse simulation employed in the present paper  and their effect on our results.

Finally, the luminosity of the whole supernova in right-handed neutrinos as a function of time is presented at Fig.\ref{luminosity}. Two curves at this figure represent the uncertainty of our result due to the ignorance of the exact conditions inside the supernova core. We checked the reliability of the employed code "Boom" by comparing its results with the results reported in \cite{Sumiyoshi}. In this paper the results of the simulations are presented for two possible equations of state (EOS) of nuclear matter. Boom results for densities and temperatures are in a good agreement with the results from \cite{Sumiyoshi} for a more stiff EOS. However, a softer EOS leads to larger densities and temperatures \cite{Sumiyoshi}, which leads to the increase of the luminosity. Moreover, the degree of deleptonization predicted by Boom is larger than such in \cite{Sumiyoshi}.
 %and in some other works (see e.g. ).
As a result, the mass of the inner core predicted by Boom is smaller than such in \cite{Sumiyoshi}. As the inner core gives the main contribution to the right-handed neutrino emission, the luminosity calculated on the basis of Boom results may occur to be underestimated.\footnote{We thank N. V. Mikheev for attracting our attention to this source of uncertainty in our calculations.} In order to take into account all the above uncertainties of the astrophysical nature, we use the following procedure. We increase the Boom densities and temperatures ``by hand'' by 60\% and 40\% correspondingly, %(compared to the Boom code output)
 keeping in the integral (\ref{emission rate}) the spherical volume elements, $d^3r=4\pi r^2 dr,$  unchanged. This  automatically increases the mass of emitting matter by 60\%.   Basing on results of \cite{Sumiyoshi}, we expect that such procedure leads to a reasonable %conservative
estimate of the uncertainty associated with the ignorance of the parameters of the supernova core. The resulting luminosity is represented by the upper curve at Fig.\ref{luminosity}.
  %corresponds to luminosity calculated with use of such for densities and temperatures increased ``by hand'' by 60\% and 40\% correspondingly (compared to the Boom code output).
 % Basing on results of \cite{Sumiyoshi}, we expect that such procedure leads to a reasonable %conservative estimate of the uncertainty associated with the ignorance of the parameters of the supernova core. We point out that the increase of the density by 60\% automatically implies the increase of the mass of the emitting matter

One can see that the luminosity grows abruptly at bounce and stays almost constant during at least 250 ms, being equal to $(0.5-1.8) \cdot10^{50}$ erg/s for $\mu_\nu= 10^{-13} \mu_B.$
Note that the luminosity does not show any signatures of decrease at $t=250$ ms -- the greatest time accessible to date in our numerical simulation. We expect the total amount of energy emitted during the collapse in right-handed neutrinos to be a factor of (2-4) greater than the energy emitted in the first 250 ms after bounce. This means, in particular, that in order to inject $10^{51}$ ergs in the supernova envelop according to the Dar's mechanism, the neutrino magnetic moment should equal $(2-6)\cdot10^{-13}\mu_B.$ This estimate is in agreement with the estimate presented in \cite{Kuznetsov:2009we}. %, namely $\mu_\nu=(3-10)\cdot10^{-13}\mu_B.$
Remind that the constraints on the Dirac magnetic moment presented in two last lines of Table \ref{bounds on mu} may be, in general, invalid if the exploding star possesses strong magnetic field, which is required for the Dar's mechanism to work.

%Therefore, in order to em This can, in particular, broaden the range of the neutrino magnetic moment values sufficient for the Dar's mechanism to work; in this case our estimate gets close to the estimate presented in \cite{Kuznetsov:2009we}, namely $\mu_\nu=(0.3-1.0)\cdot10^{-12}\mu_B.$

%This means, in particular, that in order to inject $10^{51}$ ergs in the supernova envelop in the first 250 ms of a collapse according to the Dar's mechanism, the neutrino magnetic moment should be only slightly less than $10^{-12}\mu_B$. Lower values of the neutrino magnetic moment do not produce enough energy (remind that the luminosity scales as $\mu_\nu^2$), while higher values violate the constraints from cooling rates of red giants and white dwarfs.

It should be noted that we do not take into account the back reaction of the right-handed neutrino emission on the supernova dynamics. This can be justified if the energy loss rate due to the right-handed neutrino emission is negligible compared to the total energy loss rate, which is of order of $10^{53}$ erg/s.  On can see that this is the case for $\mu_\nu\lesssim 10^{-12}\mu_B.$ In fact it is this reasoning which, along with the requirement that the energy loss due to the sterile neutrino emission should not be too high,  provides the bound in the fourth line of Table \ref{bounds on mu} \cite{Kuznetsov:2009zm} (see also \cite{Goldman:1987fg}-\cite{Nussinov:1987zr},\cite{Kuznetsov:2007mp}).

We compare our result with the results of previous studies \cite{Barbieri:1988nh}\cite{Kuznetsov:2007mp} in Table \ref{comparison}. From the first glance one may think that we merely confirm the result of \cite{Barbieri:1988nh}. However, this is not the case; the approximate coincidence between the present result and the result of Barbieri and Mohapatra  \cite{Barbieri:1988nh} is accidental. Spin flip rate used in the present work is larger than in \cite{Barbieri:1988nh} (and coincides with such in  \cite{Kuznetsov:2007mp}), while conditions inside the supernova core  employed in the present work are less favorable for right-handed neutrino emission compared to conditions employed in \cite{Barbieri:1988nh}. These two important  distinctions approximately compensate each other, which explains the apparent proximity of our result to the result of \cite{Barbieri:1988nh}.
%conditions are  we used in our calculations a larger spin flip rate  less favorable for  right-handed neutrino emission  in both the expression for the spin flip rate and the supernova core model are different in  \cite{Barbieri:1988nh} and in the present work our calculation differs from the calculation presented in \cite{Barbieri:1988nh} both by the expression for the spin flip rate and by  is underestimated, while temperature  the supernova core conditions are assumed the
On the other hand, the difference between the present result and the result of Kuznetsov and Mikheev \cite{Kuznetsov:2007mp} stems solely from the difference in the supernova core models.
%In the previous works no dynamical supernova model was employed,  the luminosity being calculated for a static sphere with some fixed typical values of supernova core volume, density, electron and neutrino chemical potentials and temperature. As the spin flip rate is rather sensitive to all these parameters, the uncertainties were rather big.
Anyhow, we manage to substantially reduce the overall uncertainty which was present in the literature.

\begin{table}[t]\label{comparison}
\begin{center}
\begin{tabular}{|l|c|}
\hline
Reference & Supernova core luminosity   \\
 & for $\mu_\nu=10^{-13}\mu_B$ in units $10^{50}$ erg/s\\
\hline
\hline
Barbieri and Mohapatra \cite{Barbieri:1988nh} & 0.4-4 \\
\hline
Kuznetsov and Mikheev \cite{Kuznetsov:2007mp} & 3.8-22 \\
\hline
present work & 0.5-1.8 \\
\hline
\end{tabular}
\end{center}
\caption{Previously reported and present results for the supernova core luminosity in right-handed neutrinos during several hundreds of milliseconds after bounce. The approximate coincidence between the present result and the result of \cite{Barbieri:1988nh} is accidental. Spin flip rate used in the present work is larger than in \cite{Barbieri:1988nh}, while conditions inside the supernova core employed in the present work are  less favorable for  right-handed neutrino emission compared to those in \cite{Barbieri:1988nh}.
%conditions are  we used in our calculations a larger spin flip rate  less favorable for  right-handed neutrino emission  in both the expression for the spin flip rate and the supernova core model are different in  \cite{Barbieri:1988nh} and in the present work our calculation differs from the calculation presented in \cite{Barbieri:1988nh} both by the expression for the spin flip rate and by  is underestimated, while temperature  the supernova core conditions are assumed the
The difference between the present result and the result of \cite{Kuznetsov:2007mp} stems solely from the difference in the supernova core models.}
\end{table}

\section{High-energy neutrino signal due to Dirac neutrino magnetic moment in water cherenkov detectors}
In order for right-handed neutrinos to reveal themselves in the detectors, they should be converted to left-handed ones due to the spin precession in the interstellar magnetic field. If this occurs, then high-energy (with a spectrum peaked at (100-150) MeV, see Fig. \ref{spectrum}) neutrinos may be registered in the detectors {\it simultaneously} with the ordinary neutrino signal (with a spectrum exponentially suppressed for energies greater than $\sim 70$ MeV). We emphasize that we do not study the ordinary supernova neutrino signal. We are interested in high-energy neutrinos only.

The number of neutrino events with energies greater than $E_0$ in a water Cherenkov detector reads
\be\label{event rate detailed}
N= \kappa \cdot \int_{E_0}^\infty dE \frac{M_{\rm H_2O}}{m_{\rm H_2O} }~\frac{\sigma(E)}{4\pi D^2}  \int dt \frac{dN_{\nu_R}}{dEdt}.
\ee
Here\\
$\kappa$  is the fraction of $\nu_{eR}$ converted to $\nu_{eL}$ in the interstellar space on the way from the supernova to the Earth,\\
$M_{\rm H_2O}$ and $m_{\rm H_2O}$ are the fiducial mass of water and the mass of the H$_2$O molecule, correspondingly,\\
$D$ is the distance from the supernova (we use a reference value $D=10$ kpc in what follows), \\
$\sigma(E)$ is the cross section of the reaction $\nu_e + \Ox \ar e^- + \F$ for the neutrino with energy $E,$ through which the electron neutrinos are registered in water detectors. We employ the cross section presented in \cite{Haxton:1987kc}.\\

Coefficient $\kappa$ deserves some special attention. Two effects should be taken into account while calculating it, the neutrino spin precession in the interstellar magnetic field and the neutrino flavor mixing. We elaborate on it in the Appendix. It is normally somewhat less than $1/2$ for $\mu_\nu\gtrsim 10^{-13} \mu_B.$ For our estimates we take it to be $0.3$ (see the Appendix).

We find that the first 250 ms after bounce should provide (3-13) high-energy events in  Super-Kamiokande (with fiducial mass equal to 22 kt) if $\mu_\nu=10^{-13}\mu_B$. This confirms the rough estimate presented in \cite{Lychkovskiy:2008da}. One can see that Super-Kamiokande is well sensitive to the Dirac neutrino magnetic moment of order of $10^{-13}\mu_B.$

\section{Spin flips of neutrinos with Majorana neutrino magnetic moment inside the inner supernova core}

If neutrinos are Majorana fermions, spin flips do not produce sterile neutrino species. Instead they convert electron neutrinos into non-electron antineutrinos. Remind that in the standard picture of the collapse electron neutrinos inside the supernova core form a highly degenerate fermion gas, while other neutrino and antineutrino species are nearly absent. The reason for this is that standard model interactions conserve lepton number and lepton flavor. Majorana neutrino spin flips violate both conservation laws. As a result, for a sufficiently large Majorana transition moments, muon and tau antineutrinos may become as numerous as electron neutrinos. In principle, this may alter both the supernova dynamics and the supernova neutrino signal. The impact of the decrease of the electron neutrino degeneracy on the shock dynamics is considered in \cite{Fuller:1987}\cite{Rampp:2002kn}. In \cite{Lychkovskiy:2008da} it is pointed out that $\nu_e\rightarrow\bar\nu_\mu,\bar\nu_\tau$ transition due to spin flips of neutrinos with Majorana magnetic moment in collisions with charged particles may change the flavor composition of the supernova neutrino output. In particular, it the composition of the neutronization burst could be altered, which would drastically increase its observability in water Cherenkov detectors\footnote{In the standard picture of the collapse only electron neutrinos constitute the neutronization burst (see e.g. \cite{Thompson:2002mw}). This complicates the possibility of observing neutronization burst in water Cherenkov detectors, as they are predominantly sensible to electron {\it antineutrinos} through the inverse beta-decay, $\bar \nu_e p\rightarrow e^+ n.$ Spin flips inside the core convert electron neutrinos to non-electron antineutrinos. Some fraction of non-electron anti-neutrinos is generically further converted to electron antineutrinos due to the MSW-effect in the supernova envelope, see e.g. \cite{Dighe:1999bi}. This may lead to a large fraction of well-detectable electron antineutrinos in the neutronization burst as it reaches the Earth. However it should be stressed that self-consistent calculations involving careful account for antineutrino transport are necessary in order to verify this effect. The similar effect was described in \cite{Ando:2002sk}-\cite{Akhmedov:2003fu}. In these works resonant spin-flavor transitions in the presence of large magnetic fields were considered instead of spin flips in collisions with charged particles.}. Finally, we note that the emergence of non-electron antineutrinos inside the core in addition to electron neutrinos should fasten the cooling rate of the core due to the neutrino diffusion, which may influence both supernova dynamics and neutrino signal.

All these effects imply that considerable fraction of electron neutrinos experience spin flips. Therefore, the back reaction of spin flips on the supernova core evolution should be taken into account in order to study these effects in detail. We do not perform such analysis in the present paper. However we make a rough estimate of the Majorana magnetic moment value necessary to equilibrate electron neutrinos with non-electron antineutrinos in the inner core at a timescale of 1 ms (this is a relevant timescale for shock propagation and neutronization burst). For the typical parameters of the supernova core matter presented in Section 2 (before eq.(\ref{Fermi energies})) we find this value to be $\mu_{\rm \nu}\simeq10^{-12}\mu_B.$ This value is not yet constrained by terrestrial experiments and astrophysical considerations. Therefore it is interesting to thoroughly explore the impact of the Majorana neutrino magnetic moment on the physics of core-collapse supernova.

\section{Conclusions}

We have performed a calculation of the luminosity of the supernova core in right-handed neutrinos in the first half of a second of a collapse, assuming that neutrinos are Dirac fermions with magnetic moment.
%from the beginning of the infall.
It is shown that the luminosity grows abruptly at bounce and stays almost constant during at least 250 ms, being equal to $(0.5-1.8)\cdot10^{50}$ erg/s for $\mu_\nu= 10^{-13} \mu_B.$ We expect that it is significant also for larger times; further work is necessary to demonstrate this explicitly. However, already obtained results allow to conclude that Super-Kamiokande may register at least few high-energy neutrino events if $\mu_\nu$ is of order of $10^{-13} \mu_B.$

Also we point out that if neutrinos are Majorana fermions with magnetic moment around $10^{-12} \mu_B,$ then their spin flips inside the inner core may equilibrate electron neutrinos with non-electron antineutrinos. This may affect the supernova dynamics and the supernova neutrino signal.

%Neutrino signal from a future nearby supernova is likely to allow to probe neutrino magnetic moment $\mu$ up to $\mu \gtrsim 10^{-13} \mu_B$ at Super-Kamiokande and $\mu \gtrsim 0.5\cdot 10^{-13} \mu_B$ at a future Mt-scale detector, provided neutrino is a Dirac fermion. The detection of high-energy neutrinos (typical energies -- (150-200) MeV) would be a smoking gun for such values of $\mu$. With the Mt-scale detector another signature of the Dirac magnetic moment may be observed: the deficit of the neutronization burst neutrinos compared to what is expected in the conventional scenario with negligible neutrino magnetic moment. If both signatures are observed, one is able to extract the value of the magnetic moment from the observations.
%
%

\section*{Acknowledgments}
The authors are grateful to M.I. Vysotsky, D.K. Nadyozhin and N.V. Mikheev for valuable comments. OL appreciates the fruitful discussion at the Theoretical Physics Department of the Yaroslavl State University.
 SB thanks K.Nomoto  and H.Murayama for hospitality at  IPMU in Tokyo; his work is supported by World Premier International Research Center Initiative (WPI), MEXT, Japan, by grants NSh-2977.2008.2, NSh-3884.2008.2, RFBR-07-02-00830-a and by a grant IB7320-110996 of the Swiss National Science
Foundation.
The work of OL is supported by the Dynasty Foundation scholarship and grants NSh-4568.2008.2, RFBR-07-02-00830-a, RFBR-08-02-00494-a.

\section*{Appendix: neutrino spin precession and flavor transformation in the interstellar space}

Here we estimate the coefficient $\kappa.$

%\subsection{Neutrino propagation in the interstellar space and $\nu_R\ar\nu_L$ transition in the galactic magnetic field}

The interaction of the neutrino magnetic moment with the magnetic field $\B$ leads to the neutrino spin precession (or, in other words, to $\nu_R \leftrightarrow \nu_L$ oscillations), which is described (in the ultra-relativistic case) by \cite{Voloshin et al}
\be \label{evolution eq}
i\frac{d}{dx}\nu(x)= (E+\mu {\boldsymbol \sigma} {\Bp(x)} )\nu(x).
\ee
Here
$ \nu (x)=\left(\begin{array}{l} \nu_L(x) \\ \nu_R(x)
\end{array}\right),$
$E$ is neutrino energy, $\boldsymbol\sigma$ is a vector constructed from Pauli matrices, and $\Bp(x)$ is a component of $\B$ normal to the neutrino momentum.
If for every $x$ magnetic field $\Bp(x)$ lies in the same plane, then the phase of oscillations is given by
\be\label{phase general}
\phi=\int  \mu B_\perp (x) dx.
\ee
The oscillation probability, $P(\nu_R\ar\nu_L)=\sin^2\phi,$ may be easily calculated for this case.  For the constant magnetic field
%, $\Bp(x)=\Bp,$
one gets
\be\label{phase}
\phi=\mu_\nu B_\perp x= 0.9 \lb \frac{\mu}{10^{-13}\mu_B} \rb \lb \frac{B_\perp }{{\rm \mu G}}\rb
\lb \frac{x}{10~{\rm kpc}} \rb.
\ee

Galactic magnetic field has a complicated structure (see, for example, \cite{Vallee}, \cite{Han:2007uu}). Its typical strength is not less than 1 $\mu$G, and probably somewhat larger. It can be represented as the sum of regular (large-scale) and random (small-scale) components. Length scales of the random component are much smaller than 1 kpc, therefore, according to (\ref{phase}), this component is irrelevant for our purposes.\footnote{
Strictly speaking, relevance of the random magnetic field to the spin rotation is determined by
$\gamma=\mu^2 \langle B^2 \rangle L_c x$
\cite{Nicolaidis:1991da},
where $L_c$ is a field variation length scale. Taking $\mu=10^{-12}\mu_B, ~x=10$ kpc and $B \sim 1~\mu$Gs, $L_c\sim 10 $ pc (see \cite{Han:2007uu} and reference therein), one obtais
$\gamma\sim 0.1\ll 1.$ This means that the effect of the random magnetic field on the spin precession may be neglected.
}
 Length scales of the regular component are of order of 1 kpc. In the galactic disk regular magnetic field is directed along the spiral arms, clockwise or counterclockwise depending on the spiral arm. There is a number of galactic magnetic field models (see the above mentioned references). We use a phenomenological model described in \cite{Vallee}. It fits well the data extracted from observations of 350 pulsars. Also it reproduces the main qualitative features of the radial dependence of the magnetic field in the inner galaxy, i.e. two field reversals at $\sim 4.5$ and $\sim 6.5$ kpc from the center of the galaxy, as well as the characteristic strength of the magnetic field.

We consider a frequently discussed case of a supernova exploding in the inner part of the disk of our galaxy, $D=10$ kpc away from the Solar system. For simplicity we assume that it is situated on the line which connects Solar system and galactic center.
% Remind that the Solar system itself is $\sim8.5$ kpc away from the galactic center.
The radial dependence of $B_\perp (r)$  reads \cite{Vallee}
\be \label{B regular}
B_\perp (r) = \bc
0.9 ~\mu{\rm G} & 0~\kpc < r \le 2 ~{\rm kpc} \\
3.8 ~\mu{\rm G} & 2~\kpc < r \le 3~ {\rm kpc} \\
3.1 ~\mu{\rm G} & 3~\kpc < r \le 4~ {\rm kpc} \\
-2.2 ~\mu{\rm G} & 4~\kpc < r \le 5~ {\rm kpc} \\
-1.9 ~\mu{\rm G} & 5~\kpc < r \le 6~ {\rm kpc} \\
1.9 ~\mu{\rm G} & 6~\kpc < r \le 7~ {\rm kpc} \\
2.5 ~\mu{\rm G} & 7~\kpc < r \le 8~  {\rm kpc,} \\
%2.2 ~\mu{\rm G} & 8~\kpc < r \le 9~ {\rm kpc}
              \ec
\ee
Here $r$ is the galactocentric distance. The distance from the Sun to the galactic center is taken to be 7.2 kpc in \cite{Vallee}.

From (\ref{phase general}) and (\ref{B regular}) one obtains the probability of $\nu_R\ar\nu_L$ oscillations:
\be\label{P}
P_{\nu_R\ar\nu_L}=\sin^2  \lb 1.1 \frac{\mu}{10^{-13}\mu_B}\rb.
\ee

 Two cases should be distinguished. For $\mu \gtrsim \mu_{th} \simeq   10^{-13}\mu_B$ the probability oscillates  rapidly with $\mu,$  phase being strongly dependent on the actual magnetic field along the line of sight to the supernova. This means, in fact, that for $\mu \gtrsim \mu_{th}$ one should consider
the phase $\phi$ as a uniformly distributed random value. In this case $P_{\nu_R\ar\nu_L}$ is also a random value. Its expectation value is
$P_{av}=0.5$, and $P_{\nu_R\ar\nu_L}> 0.025 $ with $90\%$ probability.

Alternatively, if $\mu \lesssim \mu_{th},$ sine may be approximated by its argument in eq.(\ref{P}). In this case $P_{\nu_R\ar\nu_L}$ is proportional to $\mu^2,$ magnetic field profile affecting only the coefficient of proportionality.

Although the exact value of the phase in eq.(\ref{P}) depends on the factual magnetic field profile, the above described qualitative behavior of the probability $P_{\nu_R\ar\nu_L}$ with respect to $\mu$ is common for any profile. The value of $\mu_{th}$ is expected to be of order of $10^{-13}\mu_B$ for the considered case of a supernova in the inner part of the galactic disk.
%Although the exact value of $\mu_{th}$ depends on the factual magnetic field profile, it is clearly expected to be of order %of $10^{-13}\mu_B.$ The above described qualitative behavior of the probability $P_{\nu_R\ar\nu_L}$ with respect to $\mu$
%is common for any profile.

One should also take into account flavor transformations along with the spin precession in the problem involved. Right-handed electron neutrinos, produced in a supernova, $\nu_{eR},$  quickly decohere into the mixture of $\nu_{1R}$ and $\nu_{2R}$ (see e.g. \cite{Dighe:1999bi}). The corresponding fractions in the mixture are equal to
$\cos^2\theta_{12}$ and $\sin^2\theta_{12},$ $\theta_{12}\simeq 30^o$ being the mixing angle.\footnote{ For simplicity, we assume that the vacuum mixing of right-handed neutrinos is equivalent to such for left-handed neutrinos. This allows to obtain a definite expression for the conversion coefficient $\kappa.$ However, this simplification influences only the mixing-angle prefactor in eq.(\ref{kappa}), which is in any case generically of order of unity.} In the interstellar medium  $\nu_{1R}\ar\nu_{1L}$ and $\nu_{2R}\ar\nu_{2L}$ transitions occur with the probability $P_{\nu_R\ar\nu_L}$ as described above. Finally, in the detector $\nu_{1L}$ and $\nu_{2L}$ show themselves as electron neutrinos with probabilities $\cos^2\theta_{12}$ and $\sin^2\theta_{12}$ correspondingly. Combining all the probabilities together, one finds the fraction $\kappa$ of right-handed electron neutrinos, $\nu_{eR},$  which are converted to left-handed electron neutrinos, $\nu_{eL},$ in the interstellar space on the way from the supernova to the Earth:
\be\label{kappa}
\kappa =(\cos^4\theta_{12} + \sin^4\theta_{12})P_{\nu_R\ar\nu_L}=(1-0.5\sin^2 2\theta_{12})P_{\nu_R\ar\nu_L}
\approx 0.6 P_{\nu_R\ar\nu_L}.
\ee
For $\mu \gtrsim \mu_{th}$ one finally obtains $\kappa\approx0.3$


\begin{thebibliography}{99}

  %\cite{Decamp:1989fr}
\bibitem{Decamp:1989fr}
  D.~Decamp {\it et al.}  [ALEPH Collaboration],
  ``A precise determination of the number of families with light neutrinos and of the Z boson partial widths,''
  Phys.\ Lett.\  B {\bf 235}, 399 (1990).
  %%CITATION = PHLTA,B235,399;%%

  %\cite{Barger:2003zg}
\bibitem{Barger:2003zg}
  V.~Barger, J.~P.~Kneller, H.~S.~Lee, D.~Marfatia and G.~Steigman,
  ``Effective number of neutrinos and baryon asymmetry from BBN and WMAP,''
  Phys.\ Lett.\  B {\bf 566}, 8 (2003).
  %%CITATION = PHLTA,B566,8;%%

%\cite{Lee:1977tib}
\bibitem{Lee:1977tib}
  B.~W.~Lee and R.~E.~Shrock,
  ``Natural Suppression Of Symmetry Violation In Gauge Theories: Muon - Lepton
  And Electron Lepton Number Nonconservation,''
  Phys.\ Rev.\  D {\bf 16}, 1444 (1977).
  %%CITATION = PHRVA,D16,1444;%%

%\cite{Fujikawa:1980yx}
\bibitem{Fujikawa:1980yx}
  K.~Fujikawa and R.~Shrock,
  ``The Magnetic Moment Of A Massive Neutrino And Neutrino Spin Rotation,''
  Phys.\ Rev.\ Lett.\  {\bf 45}, 963 (1980).
  %%CITATION = PRLTA,45,963;%%

%\cite{Voloshin:1987qy}
\bibitem{Voloshin:1987qy}
  M.~B.~Voloshin,
  ``On Compatibility of Small Mass with Large Magnetic Moment of Neutrino,''
  Sov.\ J.\ Nucl.\ Phys.\  {\bf 48}, 512 (1988)
  [Yad.\ Fiz.\  {\bf 48}, 804 (1988)].
  %%CITATION = YAFIA,48,804;%%

  %\cite{Vysotsky:2002yu}
\bibitem{Vysotsky:2002yu}
  M.~Vysotsky,
  ``Neutrino magnetic moment: Review talk at the workshop 'Search for dark
  matter and neutrino magnetic moment', ITEP,11.12.2001,''
  Mod.\ Phys.\ Lett.\  A {\bf 18}, 877 (2003)
  [arXiv:hep-ph/0209070].
  %%CITATION = MPLAE,A18,877;%%

%*******************************************************************************

%\cite{Cisneros:1970nq}
\bibitem{Cisneros:1970nq}
  A.~Cisneros,
  ``Effect of neutrino magnetic moment on solar neutrino observations,''
  Astrophys.\ Space Sci.\  {\bf 10}, 87 (1971).
  %%CITATION = APSSB,10,87;%%


  %\cite{Voloshin:1986ty}
\bibitem{Voloshin:1986ty}
  M.~B.~Voloshin and M.~I.~Vysotsky,
  ``Neutrino Magnetic Moment And Time Variation Of Solar Neutrino Flux,''
  Sov.\ J.\ Nucl.\ Phys.\  {\bf 44}, 544 (1986)
  [Yad.\ Fiz.\  {\bf 44}, 845 (1986)].
  %%CITATION = YAFIA,44,845;%%

    %\cite{Okun:1986uf}
\bibitem{Okun:1986uf}
  L.~B.~Okun,
  ``On The Electric Dipole Moment Of Neutrino,''
  Sov.\ J.\ Nucl.\ Phys.\  {\bf 44}, 546 (1986)
  [Yad.\ Fiz.\  {\bf 44}, 847 (1986)].
  %%CITATION = YAFIA,44,847;%%

  %\cite{Okun:1986hi}
\bibitem{Okun:1986hi}
  L.~B.~Okun, M.~B.~Voloshin and M.~I.~Vysotsky,
  ``Electromagnetic Properties Of Neutrino And Possible Semiannual Variation
  Cycle Of The Solar Neutrino Flux,''
  Sov.\ J.\ Nucl.\ Phys.\  {\bf 44}, 440 (1986)
  [Yad.\ Fiz.\  {\bf 44}, 677 (1986)].
  %%CITATION = YAFIA,44,677;%%

  %\cite{Veselov:1987ky}
\bibitem{Veselov:1987ky}
  A.~I.~Veselov, M.~I.~Vysotsky and V.~P.~Yurov, ``Solar neutrino flux half year variations from Davis' 1979 - 1982 data,''
  Sov.\ J.\ Nucl.\ Phys.\  {\bf 45}, 865 (1987)
  [Yad.\ Fiz.\  {\bf 45}, 1392 (1987)].
  %%CITATION = YAFIA,45,1392;%%

  \bibitem{Voloshin et al}
  L.~B.~Okun, M.~B.~Voloshin and M.~I.~Vysotsky,
  ``Neutrino electrodynamics and possible consequences for solar neutrinos,''
  Sov.\ Phys.\ JETP {\bf 64}, 446 (1986)
  [Zh.\ Eksp.\ Teor.\ Fiz.\  {\bf 91}, 754 (1986)].
  %%CITATION = ZETFA,91,754;%%
%**************************************************************

%\cite{Dar:1987yv}
\bibitem{Dar:1987yv}
  A.~Dar,
"Neutrino magnetic moment may solve the supernovae problem,"
 PRINT-87-0178-IAS,Princeton.

%\cite{Goldman:1987fg}
\bibitem{Goldman:1987fg}
  I.~Goldman, Y.~Aharonov, G.~Alexander and S.~Nussinov,
"Implications of the supernova SN1987A neutrino signal,"
  Phys.\ Rev.\ Lett.\  {\bf 60}, 1789 (1988).
  %%CITATION = PRLTA,60,1789;%%

%\cite{Lattimer:1988mf}
\bibitem{Lattimer:1988mf}
  J.~M.~Lattimer and J.~Cooperstein,
  ``Limits on the Neutrino Magnetic Moment from SN 1987a,''
  Phys.\ Rev.\ Lett.\  {\bf 61}, 23 (1988).
  %%CITATION = PRLTA,61,23;%%

%\cite{Barbieri:1988nh}
\bibitem{Barbieri:1988nh}
  R.~Barbieri and R.~N.~Mohapatra,
  ``Limit on the Magnetic Moment of the Neutrino from Supernova SN 1987a
  Observations,''
  Phys.\ Rev.\ Lett.\  {\bf 61}, 27 (1988).
  %%CITATION = PRLTA,61,27;%%

%\cite{Notzold:1988kz}
\bibitem{Notzold:1988kz}
  D.~Notzold,
  %``NEW BOUNDS ON NEUTRINO MAGNETIC MOMENTS FROM STELLAR COLLAPSE,''
"New bounds on neutrino magnetic moments from stellar collapse,"
  Phys.\ Rev.\  D {\bf 38}, 1658 (1988).
  %%CITATION = PHRVA,D38,1658;%%



%\cite{Nussinov:1987zr}
\bibitem{Nussinov:1987zr}
  S.~Nussinov and Y.~Rephaeli,
  ``Magnetic Moments Of Neutrinos: Particle And Astrophysical Aspects,''
  Phys.\ Rev.\  D {\bf 36}, 2278 (1987).
  %%CITATION = PHRVA,D36,2278;%%

%*********************************************************************

%\cite{Voloshin:1988xu}
\bibitem{Voloshin:1988xu}
  M.~B.~Voloshin,
"Resonant helicity flip of electron-neutrino magnetic moment and dynamics of supernova,"
  Phys.\ Lett.\  B {\bf 209}, 360 (1988).
  %%CITATION = PHLTA,B209,360;%%

%\cite{Voloshin:1988xs}
\bibitem{Voloshin:1988xs}
  M.~B.~Voloshin,
  "On dynamics of electron neutrino inside supernova and bounds on magnetic moment of electron neutrino,"
  JETP Lett.\  {\bf 47}, 501 (1988)
  [Pisma Zh.\ Eksp.\ Teor.\ Fiz.\  {\bf 47}, 421 (1988)].
  %%CITATION = ZFPRA,47,421;%%

%\cite{Okun:1988qs}
\bibitem{Okun:1988qs}
  L.~B.~Okun,
  ``$\nu_e- \nu_e$ scattering and the possibility of a resonance change of  neutrino
  helicity in the magnetic field of a supernova,''
  Sov.\ J.\ Nucl.\ Phys.\  {\bf 48}, 967 (1988).
  %%CITATION = SJNCA,48,967;%%

%\cite{Blinnikov:1988xq}
\bibitem{Blinnikov:1988xq}
  S.~I.~Blinnikov and L.~B.~Okun,
"Supernova models and the neutrino magnetic moment,"
   Pis'ma Astron. Zh. {\bf 14}, 867 (1988).
  %%CITATION = ITEP-88-123;%%

%****************************************************************************

%\cite{Bernstein:1963qh}
\bibitem{Bernstein:1963qh}
  J.~Bernstein, M.~Ruderman and G.~Feinberg,
  ``Electromagnetic Properties of the neutrino,''
  Phys.\ Rev.\  {\bf 132}, 1227 (1963).
  %%CITATION = PHRVA,132,1227;%%

  %\cite{Fukugita:1987uy}
\bibitem{Fukugita:1987uy}
  M.~Fukugita and S.~Yazaki,
  ``Reexamination of Astrophysical and Cosmological Constraints on the Magnetic
  Moment of Neutrinos,''
  Phys.\ Rev.\  D {\bf 36}, 3817 (1987).
  %%CITATION = PHRVA,D36,3817;%%


 \bibitem{Blinnikov&Dunina}
  S.I. Blinnikov, N.V. Dunina-Barkovskaya,
The cooling of hot white dwarfs: a theory with non-standard
weak interactions and a comparison with observations,
MNRAS {\bf 266} 289 (1994)


%\cite{Raffelt:1999gv}
\bibitem{Raffelt:1999gv}
  G.~G.~Raffelt,
  ``Limits on neutrino electromagnetic properties: An update,''
  Phys.\ Rept.\  {\bf 320}, 319 (1999).
  %%CITATION = PRPLC,320,319;%%

%****************************************************************************

%\cite{Arpesella:2008mt}
\bibitem{Arpesella:2008mt}
  C.~Arpesella {\it et al.}  [The Borexino Collaboration],
  ``Direct Measurement of the Be-7 Solar Neutrino Flux with 192 Days of
  Borexino Data,''
  Phys.\ Rev.\ Lett.\  {\bf 101}, 091302 (2008)
  [arXiv:0805.3843 [astro-ph]].
  %%CITATION = PRLTA,101,091302;%%

%\cite{Beda:2007hf}
\bibitem{Beda:2007hf}
  A.~G.~Beda {\it et al.},
  ``The first result of the neutrino magnetic moment measurement in the   GEMMA
  experiment,'' Yad.Phys. {\bf 70}, 1925 (2007), arXiv:0705.4576 [hep-ex].
  %%CITATION = ARXIV:0705.4576;%%

  %\cite{Daraktchieva:2003dr}
\bibitem{Daraktchieva:2003dr}
  Z.~Daraktchieva {\it et al.}  [MUNU Collaboration],
  ``Limits on the neutrino magnetic moment from the MUNU experiment,''
  Phys.\ Lett.\  B {\bf 564}, 190 (2003)
  [arXiv:hep-ex/0304011].
  %%CITATION = PHLTA,B564,190;%%


%*****************************************************************************

%\cite{Kuznetsov:2009zm}
\bibitem{Kuznetsov:2009zm}
  A.~V.~Kuznetsov, N.~V.~Mikheev and A.~A.~Okrugin,
  ``Reexamination of a Bound on the Dirac Neutrino Magnetic Moment from the
  Supernova Neutrino Luminosity,''
  arXiv:0907.2905 [hep-ph].
  %%CITATION = ARXIV:0907.2905;%%


%\cite{Kuznetsov:2007mp}
\bibitem{Kuznetsov:2007mp}
  A.~V.~Kuznetsov and N.~V.~Mikheev,
  ``New bounds on the neutrino magnetic moment from the plasma induced neutrino
  chirality flip in a supernova,''
  JCAP {\bf 0711}, 031 (2007)
  [arXiv:0709.0110 [hep-ph]].
  %%CITATION = JCAPA,0711,031;%%

%\cite{Elmfors:1997tt}
\bibitem{Elmfors:1997tt}
  P.~Elmfors, K.~Enqvist, G.~Raffelt and G.~Sigl,
  %``Neutrinos with Magnetic Moment: Depolarization Rate in Plasma,''
  Nucl.\ Phys.\  B {\bf 503}, 3 (1997)
  [arXiv:hep-ph/9703214].
  %%CITATION = NUPHA,B503,3;%%

%\cite{Ayala:1998qz}
\bibitem{Ayala:1998qz}
  A.~Ayala, J.~C.~D'Olivo and M.~Torres,
  ``Neutrino chirality flip through photon Landau damping in supernovae,''
  Phys.\ Rev.\  D {\bf 59}, 111901 (1999)
  [arXiv:hep-ph/9804230].
  %%CITATION = PHRVA,D59,111901;%%

  %\cite{Lychkovskiy:2008da}
\bibitem{Lychkovskiy:2008da}
  O.~Lychkovskiy,
  ``Neutrino magnetic moment signatures in the supernova neutrino signal,'' proceedings of the 51th Scientific Conferense of MIPT, part II, Moscow, MIPT publishing: 2008, p.90 (in Russian); arXiv:0804.1005 [hep-ph].
  %%CITATION = ARXIV:0804.1005;%%


%***************************************************************************

%\cite{Bethe:1990mw}
\bibitem{Bethe:1990mw}
  H.~A.~Bethe,
  ``Supernova mechanisms,''
  Rev.\ Mod.\ Phys.\  {\bf 62}, 801 (1990).
  %%CITATION = RMPHA,62,801;%%
  
  %\cite{Burrows:1992ec}
\bibitem{Burrows:1992ec}
  A.~Burrows, R.~Gandhi and M.~S.~Turner,
  ``Massive Dirac neutrinos and SN1987A,''
  Phys.\ Rev.\ Lett.\  {\bf 68}, 3834 (1992).
  %%CITATION = PRLTA,68,3834;%%




%*****************************************************************************

\bibitem{Boom}
Joseph Chen-Yu Wang, A one-dimentional model of convenction in iron core collapse supernova, PhD dissertation at Uneversity of Texas at Austin, unpublished.
See also the home page for Boom code, http://en.wikiversity.org/wiki/BoomCode

\bibitem{Blinnikov et al}
S.I. Blinnikov, N.V. Dunina-Barkovskaya, and D.K. Nadyozhin, Equations of state of a Fermi gas: approximations for various degrees of relativism and degeneracy,  Astrophys.\ J.\ Supp.\ Ser. {\bf 106}, 171 (1996)

 \bibitem{Sumiyoshi}
  K. ~ Sumiyoshi {\it et al.},
  ``Postbounce evolution of core-collapse supernova: long-term effects of equation of state,''
  Astrophys.J. {\bf 629}, 922 (2005)
  [	arXiv:astro-ph/0506620v1].
  %%CITATION = PHRVA,D71,112005;%%

%**************************************************************************

%\cite{Kuznetsov:2009we}
\bibitem{Kuznetsov:2009we}
  A.~V.~Kuznetsov, N.~V.~Mikheev and A.~A.~Okrugin,
  ``Dirac-Neutrino Magnetic Moment and the Dynamics of a Supernova Explosion,''
  JETP Lett.\  {\bf 89}, 97 (2009)
  [arXiv:0903.2321 [hep-ph]].
  %%CITATION = JTPLA,89,97;%%

  %**************************************************************************

  %\cite{Haxton:1987kc}
\bibitem{Haxton:1987kc}
  W.~C.~Haxton,
  ``The Nuclear Response of Water Cherenkov Detectors to Supernova and Solar
  Neutrinos,''
  Phys.\ Rev.\  D {\bf 36}, 2283 (1987).
  %%CITATION = PHRVA,D36,2283;%%




%******************************************************************************

  %\cite{Goodman:1993zza}
\bibitem{Fuller:1987}
G.~A.~Fuller, R.~W.~ Mayle, J.~R.~Wilson, and D.~N.~Schramm,
``Resonant neutrino oscillations and stellar collapse'',
  Astrophys.\ J.\  {\bf 322}, 795 (1987).
  %%CITATION = ASJOA,406,528;%%

%\cite{Rampp:2002kn}
\bibitem{Rampp:2002kn}
  M.~Rampp, R.~Buras, H.~T.~Janka and G.~Raffelt,
  ``Core-collapse supernova simulations: Variations of the input physics,''
  arXiv:astro-ph/0203493.
  %%CITATION = ASTRO-PH/0203493;%%

  %******************************************************************************

  %\cite{Thompson:2002mw}
\bibitem{Thompson:2002mw}
  T.~A.~Thompson, A.~Burrows and P.~A.~Pinto,
  ``Shock breakout in core-collapse supernovae and its neutrino signature,''
  Astrophys.\ J.\  {\bf 592}, 434 (2003)
  [arXiv:astro-ph/0211194].
  %%CITATION = ASJOA,592,434;%%

    %\cite{Dighe:1999bi}
\bibitem{Dighe:1999bi}
  A.~S.~Dighe and A.~Y.~Smirnov,
  ``Identifying the neutrino mass spectrum from the neutrino burst from a
  supernova,''
  Phys.\ Rev.\  D {\bf 62}, 033007 (2000)
  [arXiv:hep-ph/9907423].
  %%CITATION = PHRVA,D62,033007;%%


%*******************************************************************

%\cite{Ando:2002sk}
\bibitem{Ando:2002sk}
  S.~Ando and K.~Sato,
  ``Three-generation study of neutrino spin-flavor conversion in supernova and
  implication for neutrino magnetic moment,''
  Phys.\ Rev.\  D {\bf 67}, 023004 (2003)
  [arXiv:hep-ph/0211053].
  %%CITATION = PHRVA,D67,023004;%%

%\cite{Ahriche:2003wt}
\bibitem{Ahriche:2003wt}
  A.~Ahriche and J.~Mimouni,
  ``Supernova neutrino spectrum with matter and spin flavor precession
  effects,''
  JCAP {\bf 0311}, 004 (2003)
  [arXiv:astro-ph/0306433].
  %%CITATION = JCAPA,0311,004;%%

%\cite{Ando:2003is}
\bibitem{Ando:2003is}
  S.~Ando and K.~Sato,
  ``A comprehensive study of neutrino spin-flavor conversion in supernovae  and
  the neutrino mass hierarchy,''
  JCAP {\bf 0310}, 001 (2003)
  [arXiv:hep-ph/0309060].
  %%CITATION = JCAPA,0310,001;%%

%\cite{Akhmedov:2003fu}
\bibitem{Akhmedov:2003fu}
  E.~K.~Akhmedov and T.~Fukuyama,
  ``Supernova prompt neutronization neutrinos and neutrino magnetic  moments,''
  JCAP {\bf 0312}, 007 (2003)
  [arXiv:hep-ph/0310119].
  %%CITATION = JCAPA,0312,007;%%

%*******************************************************************



\bibitem{Vallee}
  J.~P.~Vallee,
  "Pulsar-based Galactic Magnetic Map: A Large-Scale Clockwise Magnetic Field with an Anticlockwise Annulus,"
  Astrophys.\ J.\  {\bf 619}, 297 (2005).

%\cite{Han:2007uu}
\bibitem{Han:2007uu}
  J.~Han,
  ``Magnetic fields of our Galaxy on large and small scales,'' IAU Symposium {\bf 242} 55 (2008)
  arXiv:0705.4175 [astro-ph].
  %%CITATION = ARXIV:0705.4175;%%

%\cite{Nicolaidis:1991da}
\bibitem{Nicolaidis:1991da}
  A.~Nicolaidis,
  ``Random magnetic fields in the sun and solar neutrinos,''
  Phys.\ Lett.\  B {\bf 262}, 303 (1991).
  %%CITATION = PHLTA,B262,303;%%







\end{thebibliography}
\end{document}